\documentclass{article}

\usepackage{PRIMEarxiv}

\usepackage[utf8]{inputenc} 
\usepackage[T1]{fontenc}    
\usepackage{hyperref}       
\usepackage{url}            
\usepackage{booktabs}       
\usepackage{amsfonts}       
\usepackage{nicefrac}       
\usepackage{microtype}      
\usepackage{lipsum}
\usepackage{fancyhdr}       
\usepackage{graphicx}       
\graphicspath{{media/}}     
\usepackage{natbib}  
\usepackage{threeparttable} 
\usepackage{subcaption}
\usepackage{multirow}
\usepackage{lipsum} 
\usepackage{amssymb}
\usepackage{pifont}
\newcommand{\cmark}{\ding{51}}%
\newcommand{\xmark}{\ding{55}}%
\usepackage{booktabs} 
\usepackage{threeparttable} 
\usepackage{bbold}
\usepackage{amsmath}
\usepackage{array}
\usepackage{tabularx}
\usepackage{times}  
\usepackage{helvet}  
\usepackage{courier}  
\usepackage{graphicx} 

\title{Towards Long-term Annotators: A Supervised Label Aggregation Baseline
}

\author{
  Haoyu Liu\textsuperscript{$\dagger$,}\thanks{Equal contribution}~~, Fei Wang\textsuperscript{$\ddagger$,$\ast$}, Minmin Lin\textsuperscript{$\dagger$}, Runze Wu\textsuperscript{$\dagger$}, Renyu Zhu\textsuperscript{$\dagger$}, \\
  \textbf{Shiwei Zhao\textsuperscript{$\dagger$}, Kai Wang\textsuperscript{$\dagger$}, Tangjie Lv\textsuperscript{$\dagger$}, Changjie Fan\textsuperscript{$\dagger$}}\\
  \textsuperscript{$\dagger$}NetEase~Fuxi~AI~Lab \\
  \textsuperscript{$\ddagger$}University of Science and Technology of China\\
  \texttt{liuhaoyu03@corp.netease.com}
}

\begin{document}
\maketitle

\begin{abstract}
Relying on crowdsourced workers, data crowdsourcing platforms are able to efficiently provide vast amounts of labeled data. Due to the variability in the annotation quality of crowd workers, modern techniques resort to redundant annotations and subsequent label aggregation to infer true labels. However, these methods require model updating during the inference, posing challenges in real-world implementation. Meanwhile, in recent years, many data labeling tasks have begun to require skilled and experienced annotators, leading to an increasing demand for long-term annotators. These annotators could leave substantial historical annotation records on the crowdsourcing platforms, which can benefit label aggregation, but are ignored by previous works. Hereby, in this paper, we propose a novel label aggregation technique, which does not need any model updating during inference and can extensively explore the historical annotation records. We call it \textbf{SuperLA}, a \underline{\textbf{Super}}vised \underline{\textbf{L}}abel \underline{\textbf{A}}ggregation method. 
Inside this model, we design three types of input features and a straightforward neural network structure to merge all the information together and subsequently produce aggregated labels. Based on comparison experiments conducted on 22 public datasets and 11 baseline methods, we find that SuperLA not only outperforms all those baselines in inference performance but also offers significant advantages in terms of efficiency.

\end{abstract}


\section{Introduction}
Data with appropriate labels are critical for nowadays applications in artificial intelligence (AI). A main annotation source is data crowdsourcing platforms, where freelancers are employed as temporary workers and do data labeling tasks on a per-task basis. This makes it difficult to provide proper employment protection or technical training for these workers. More importantly, since the platforms commonly do not have a well-defined user persona system for these short-term annotators, it poses significant difficulties in ensuring the quality of annotations. 
On the other hand, with the rapid advancement of AI, the required data annotation has evolved from simple annotation tasks to much more complex ones recently. For instance, from tasks like classifying cats and dogs~\cite{deng2009imagenet} to classifying whether a sentence adheres to ethical and moral standards~\cite{ouyang2022training}. This results in a higher demand for skilled long-term annotators rather than random ordinary crowdsourced annotators. Therefore, label aggregation methods for long-term annotators with well-established persona systems are desired.

\begin{table}[t!]
\centering
 \begin{threeparttable}
    \begin{tabular}{lcc}
      \toprule
      \textbf{Methods} & \textbf{w/o}\textsuperscript{*} \\\midrule
    \textbf{MV} & \cmark  \\
    \textbf{All 14 works in}~\cite{zheng2017truth} & \xmark \\
    \textbf{INQUIRE}~\cite{feng2014incremental} & \xmark \\
    \textbf{LAA}~\cite{li2017aggregating} & \xmark  \\
    \textbf{SBIC}~\cite{manino2019streaming}& \xmark\\
    \textbf{BiLA}~\cite{hong2021online} & \xmark \\
    \textbf{LA}~\cite{yang2022light} & \xmark \\
    \textbf{SuperLA} (ours) & \cmark \\
     
      \bottomrule
    \end{tabular}
    \begin{tablenotes}
      \item *w/o: Do not need model updating during inference.
    \end{tablenotes}
  \end{threeparttable}
\caption{Label aggregation methods.}
\label{tab:methods_comparison}
\end{table}
However, previous label aggregation techniques mainly focus on short-term annotators, leading to inherently \textit{unsupervised} methods. For each batch of tasks, these techniques aim to iteratively estimate the hidden characteristics of annotators or tasks via maximizing likelihood estimation~\cite{dawid1979maximum,whitehill2009whose,li2017aggregating}, and subsequently deliver task truths. As shown in Table~\ref{tab:methods_comparison}, due to this iterate execution process, all these previous works require updating the model during the inference. This feature on the one hand does not fully leverage the historical records of the related annotators and on the other hand brings out a high inference time overhead since the optimization procedure is required during inference. Even though some of them like INQUIRE~\cite{feng2014incremental}, SBIC~\cite{manino2019streaming}, BiLA~\cite{hong2021online}, and LA~\cite{yang2022light} noticed this problem, they still require model updating during label aggregation, while in an incremental way. To the best of our knowledge, till now, there is no label aggregation literature that clearly splits the dataset into historical and pending data and aggregates labels in pending tasks without model updating.


The above desideratum for separating model optimization and inference as well as the leveraging of historical information inspires us to consider conducting label aggregation in a \textit{supervised} manner.
Rather than deploying previous unsupervised methods with indeterminate iteration times, supervised methods could be much more convenient and efficient in real-world implementations. 
We hereby propose SuperLA, a supervised label aggregation method. Since the under-explored nature of this direction, the challenges are mainly basic and important: what historical information and what structure should the model use. In a typical label aggregation scenario with long-term annotators, we can obtain historical records, including annotator IDs, task IDs, task truths, and annotator's annotations on tasks. To provide a consistent setting background with previous label aggregation works, we only extract features from these information and do not use any other annotator features like education status, locations, ages, and so on. For the model, how to well represent the relations between the annotators and tasks and how to merge features of different levels should be considered. More importantly, different from the previous two neural network based label aggregation models LAA~\cite{li2017aggregating} and BiLA~\cite{hong2021online}, which are trained to generatively recover the original annotations and need to train on the testing set before doing inference, we aim to devise a label aggregation model with an end-to-end inference way, i.e., model directly outputs aggregated labels and is supervised by historical task labels during training.





In light of these issues, we make such a design. Firstly, in the training set, we calculate historical accuracy of each annotator as the numerical feature and take annotator IDs as sparse features. Besides original IDs, we also convert them to multi-hot ID features so that a cross relationship could be well explored. A simple data augmentation strategy is employed, which simply mixes the order of the annotations within each task to enrich representations and improve generality. For the model part, inspired by the mature ID embedding learning model wide \& deep~\cite{cheng2016wide}, we devise a similar structure with carefully designed mechanisms to represent annotations of different labels. The learned model can be directly used for inference without any further adjustment, which is required by the previous works. Thorough experiments are conducted based on 22 datasets with 11 baselines. SuperLA substantially outperforms all other methods with lower time overhead. In a nutshell, our contributions can be summarized as follows:
\begin{itemize}
    \item We come up with an important but under explored supervised label aggregation scenario for long-term annotators. This leads to a different technical direction compared with the previous works.
    
    \item We provide simple but effective feature design and model structure to deal with supervised label aggregation problems, which could be convenient for real-world deployment. The proposed method is not only superior in terms of inference performance but also has advantages in efficiency and scalability.
    
    \item To well form the foundation of this supervised label aggregation, we conduct thorough experiments and analysis to demonstrate the superiority of the proposed method and provide a strong baseline for future research.
\end{itemize}

\section{Related Work}

Label aggregation is to predict true labels based on redundant annotations from different annotators on the same tasks~\cite{zheng2017truth}. Previous works handle each batch of tasks independently: they optimize likelihood or evidence lower bound based on the annotations of pending tasks and simultaneously deliver truths~\cite{zheng2017truth,sheng2019machine}. To enhance inference performance, the focus was on designing better inherent annotator or task models. The first and very fundamental work is DS~\cite{dawid1979maximum}, which uses confusion matrices to represent the error rates of annotators and subsequently derives truths based on these matrices and the collected annotations. Following works, including GLAD~\cite{whitehill2009whose}, LFC~\cite{raykar2010learning}, KOS~\cite{karger2011iterative}, ZC~\cite{demartini2012zencrowd}, 
Minimax~\cite{zhou2012learning}, BP~\cite{liu2012variational}, BCC~\cite{kim2012bayesian}, CBCC~\cite{venanzi2014community}, CATD~\cite{li2014confidence}, PM~\cite{aydin2014crowdsourcing},  LAA~\cite{li2017aggregating} and LA~\cite{yang2022light}, all have a very similar design strategy to DS. The difference is that, they have diverse hidden models for annotator ability or task difficulty regarding different assumptions on annotator annotation styles or task distributions.

However, all the above methods suffer time complexity issues since they need to solve optimization problems during inference. Meanwhile, in real-world crowdsourcing platforms, multiple time efficient functions like task assignment and convergence monitoring need frequent access to the temporal inference results~\cite{zheng2015qasca}, so that previous methods can not fully satisfy the real-world requirements. INQUIRE~\cite{feng2014incremental} first addresses this problem via a trivial annotator modeling strategy and a straightforward weighted vote question model to incrementally infer truths and update models. SBIC~\cite{manino2019streaming} proposes a Bayesian method with a one-pass model updating design based on Beta distribution assumption of annotator accuracy, which leads to a theoretical claim that the model is converging to the ground truth when observing a large enough number of redundant labels. However, this method does not show superiority in terms of aggregation accuracy when compared with even majority voting method on the public datasets. Recent progress is the method BiLA~\cite{hong2021online}, which currently obtains state-of-the-art performance. However, BiLA is very sensitive to the size of the pending tasks and is restricted to the full answer scenario, i.e., for a pending task, it only uses annotations that are exactly available at the current time step to infer the truth and the inferred results can not be updated when there exist other annotations in the future time steps.

\begin{figure*}[htb!]
\centering
\includegraphics[width=0.99\textwidth]{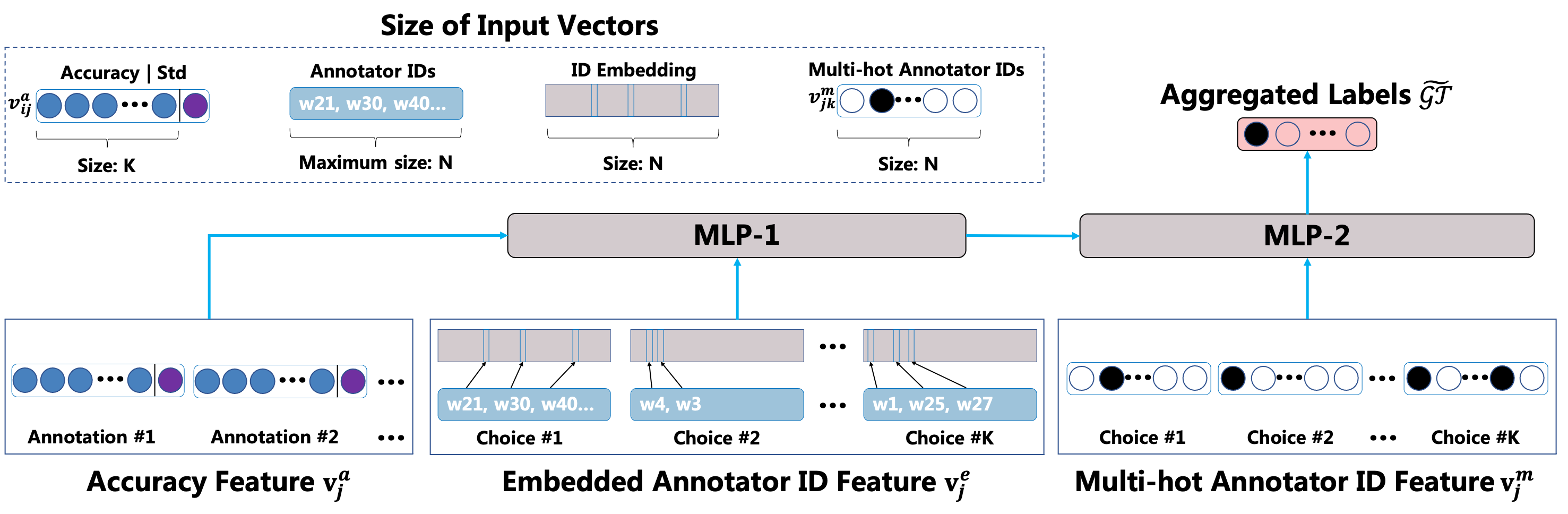} 
\caption{Framework.}
\label{fig:framework}
\end{figure*}
Moreover, all of the above methods require model updating during the inference. Even though INQUIRE, SBIC and BiLA are designed for online tasks, they are doing online inference in an incremental way rather than separately doing training and testing. Especially, BiLA needs uncertain iterations to meet convergence criteria during incremental inference, which clearly poses risks when applied in real-world applications. 

Therefore, both considering the above issues and the useful information of long-term annotators in nowadays crowdsourcing platforms, it is desired to develop supervised label aggregation methods that only do optimization on the training set and do not need any further model updating during inference on the testing set. We hereby propose SuperLA to form an exploration in this direction.

\section{Method}
The overall framework and related features are presented in Fig.~\ref{fig:framework}. 
\subsection{Problem Definition}
Given tasks $\mathcal{T}=\{\mathcal{T}_h, \mathcal{T}_p | \mathcal{T}_* = (t_1^*, t_2^*,...)\}$, where $|\mathcal{T}_h| = M_h$ denotes the historical tasks and $|\mathcal{T}_p| = M_p$ is the pending tasks. There are $N$ annotators $\mathcal{A}=(a_1, a_2,...,a_N)$, each annotator can provide a maximum of one annotation for each task, i.e., $l_{ij}\in \{\mathcal{K}|\mathcal{K}=(1, 2,...,K)\}$, representing the label that annotator $a_i$ gives to task $t_j$, the label can only be one of $K$ choices. All annotations are represented as $\mathcal{L}=\{\mathcal{L}_h, \mathcal{L}_p\}$, where $\mathcal{L}_h$ for historical tasks and $\mathcal{L}_p$ for pending tasks. For each task $t_j$, there is one truth $gt_j$ represents the real label of this task. All truths for tasks $\mathcal{T}_h$ and $\mathcal{T}_p$ are $\mathcal{GT}_h$ and $\mathcal{GT}_p$ respectively. Based on these notations, the input and output of our problem can be defined as:\\
\textbf{Input:} All task IDs $\mathcal{T}$ and annotator IDs $\mathcal{A}$, all annotations $\mathcal{L}$, historical truths $\mathcal{GT}_h$.\\
\textbf{Output:} Predicted truths for pending tasks $\widetilde{\mathcal{GT}_p}$.

\subsection{SuperLA}

From a production crowdsourcing platform, many annotator and task information like annotator education status, ages, task fields, and task pre-estimated difficulty levels are available to form abundant data features. To make a fair comparison with previous works, we exclude the above information and solely leverage the same information as previous works do, i.e., only annotation records and ID information are employed. From these information, we can first prepare three kinds of inputs before the automatic feature engineering through a neural network model.

\subsubsection{Accuracy Feature}
Historical annotator accuracy indicates the possibility that an annotator can give correct answers to tasks and could be helpful for inference in the new tasks. In the input information, there are $\mathcal{L}_h$ and $\mathcal{GT}_h$, from which we can calculate accuracy of each annotator $a_i$:
\begin{equation*}
acc_i =  \frac{\Sigma_{j=1}^{M_h} \mathbb{1}(l_{ij}=gt_{ij})}{L_{ih}},
\end{equation*}
where $L_{ih} = \Sigma_{j=1}^{M_h} \mathbb{1}(l_{ij})$, 
represents the number of historical tasks that annotator $a_i$ has answered. 
Meanwhile, the standard deviation for correct and wrong answers of an annotator $a_i$ can signal the stability of the corresponding accuracy, we calculate it by:
\begin{equation*}
std_i = \sqrt{\frac{\Sigma_{j=1}^{M_h} (\mathbb{1}(l_{ij}=gt_{ij}) - acc_i)^2}{L_{ih}}}.
\end{equation*}
Using $acc_i$ and $std_i$, for each annotation $l_{ij}$, we generate a vector $v_{ij}^a$ representing accuracy related information. The length of $v_{ij}^a$ is $K+1$, its index starts from $1$ and ends at $K+1$. At the position of $l_{ij}$, we set its value as $acc_i$, i.e., $v_{ij}^a[l_{ij}]=acc_i$. For other positions except the last one, we set their values as $\frac{1-acc_i}{K-1}$. For the last position, the value is set to $std_i$. Note that for each task, there are multiple answers from different annotators, so that we have an answer set $\mathbf{L}_{j}= \{l_{ij}|i\in(1,...,N)\}$ for task $t_j^h$. We convert all elements of this set based on the above process and get $\mathbf{v}_{j}^a=\{v_{ij}^a|i\in(1,...,N)\}$, which constitutes the accuracy feature shown in Fig.~\ref{fig:framework}. Since that different tasks could have different number of annotations, the length of $\mathbf{v}_{j}^a$ is set to be $(K+1) \times L_{max}$, where $L_{max}$ is the maximum number of annotations for one task. For those tasks with fewer annotations, we pad $0$ after their effective values.

\subsubsection{Embedded Annotator ID Feature}
\label{sec:embedded_annotator_ID_feature}
ID embedding is a mature technique to improve model generality on possible unseen feature combinations~\cite{cheng2016wide}. We employ this technique to prepare dense representations of annotators so that the input features can be enriched. As shown in Fig.~\ref{fig:framework}, for each task $t_j^h$, annotator IDs are separated into $K$ sets regarding different choices. An embedding layer is utilized after each ID set to convert the sparse ID feature to low dimension continuous feature, and the embedded annotator ID feature is hereby obtained. We denote this feature as $\mathbf{v}_{j}^e=\{v_{jk}^e|k\in(1,...,K)\}$.

\subsubsection{Multi-hot Annotator ID Feature}
The previous two features are designed as dense representations while the sparse interactions between annotators and choices should still be kept so that the frequent co-occurrences can be explored. Similar to the previous step, we first divide annotator IDs into $K$ sets for $K$ choices. Then, within each set, multi-hot encoding is conducted, where if the annotator is in the set, the corresponding value will be set to 1, otherwise 0. A visualization of this feature is also shown in Fig.~\ref{fig:framework}. We denote this feature as $\mathbf{v}_{j}^m=\{v_{jk}^m|k\in(1,...,K)\}$.

\subsubsection{Hidden Layers}
Using $\mathbf{v}_{j}^a$, $\mathbf{v}_{j}^e$ and $\mathbf{v}_{j}^m$ as the input features, our model pipeline is very straightforward and can merge these features together to deliver the final aggregated labels.

$\mathbf{v}_{j}^a$ and $\mathbf{v}_{j}^e$ are firstly forwarded into the same multi-layer perception (MLP):
\begin{equation*}
\mathbf{h}_1 = \text{MLP-1}(\text{concat}(\mathbf{v}_{j}^a, \mathbf{v}_{j}^e)).
\end{equation*}
$\mathbf{h}_1$ captures the extracted patterns from accuracy features and embedding features. Then, we concatenate $\mathbf{h}_1$ and $\mathbf{v}_{j}^m$ toghter and forward them to MLP-2:
\begin{equation*}
\mathbf{h}_2 = \text{MLP-2}(\text{concat}(\mathbf{h}_1, \mathbf{v}_{j}^m)).
\end{equation*}
The aggregated label is obtained based on the index with maximum output value in $\mathbf{h}_2$:
\begin{equation*}
\widetilde{\mathcal{GT}} = \text{argmax}(\mathbf{h}_2).
\end{equation*}
This output is further supervised by the crossentropy loss between $\widetilde{\mathcal{GT}}$ and $\widetilde{\mathcal{GT}_h}$. L2 regularizer is also adopted to mitigate potential overfitting issues.

\subsubsection{Data Augmentation} Since the time dimension is not considered in SuperLA, we can also enlarge data representations via a data augmentation strategy. In accuracy feature, we shuffle the order of each annotation vector $v_{ij}^a$ to form new input data. Such shuffle operations can be repeated many times until all the combinations are iterated. While we usually use a small replication time.

The overall pipeline is shown in Fig.~\ref{fig:framework}. The application of the model is divided into two steps. Firstly, the model is trained on historical data, i.e., $\mathcal{T}_h$, $\mathcal{L}_h$ and $\mathcal{GT}_h$. Then, during the inference, the trained model consumes $\mathcal{T}_p$, $\mathcal{L}_p$ and outputs $\widetilde{\mathcal{GT}_p}$ directly without further training. Note that, this procedure is very different from all previous label aggregation methods since they need to do model updating during the inference step.

\section{Experiments}
We conduct extensive experiments in this section to show the superiority of SuperLA in various perspectives. 
\subsection{Experimental Setup}
\subsubsection{Datasets}

\begin{table}[htbp]
\centering
\small
\begin{tabular}{lcccccc}
\hline\hline
Datasets & $|\mathcal{T}|$ & $|\mathcal{A}|$ & $|\mathcal{GT}|$ & $|\mathcal{K}|$ & $|\mathcal{L}|$  \\\hline\hline  
adult & 11040 & 825 & 333 & 4 & 89948 \\
bird & 108 & 39 & 108 & 2 & 4212 \\
cf & 300 & 461 & 300 & 5 & 1720 \\
cf\_amt & 300 & 110 & 300 & 5 & 6025 \\
dog & 807 & 109 & 807 & 4 & 8070 \\
duck & 108 & 39 & 108 & 2 & 4212 \\
face & 584 & 27 & 584 & 4 & 5242 \\
fact & 42624 & 57 & 576 & 3 & 214960 \\
ms & 700 & 44 & 700 & 10 & 2945 \\
product & 8315 & 176 & 8315 & 2 & 24945 \\
relevance & 20232 & 766 & 4460 & 4 & 97164 \\
rte & 800 & 164 & 800 & 2 & 8000 \\
sentiment & 98980 & 1960 & 1000 & 5 & 569282 \\
smile & 2134 & 64 & 159 & 2 & 19287 \\
sp & 4999 & 203 & 4999 & 2 & 27746 \\
sp\_amt & 500 & 143 & 500 & 2 & 10000 \\
trec & 19033 & 762 & 2275 & 2 & 88385 \\
tweet & 1000 & 85 & 1000 & 2 & 20000 \\
web & 2665 & 177 & 2653 & 5 & 15567 \\
zencrowd\_all & 2040 & 78 & 2040 & 2 & 20372 \\
zencrowd\_in & 2040 & 25 & 2040 & 2 & 10626 \\
zencrowd\_us & 2040 & 74 & 2040 & 2 & 11271 \\
\hline\hline
\end{tabular}
\caption{Dataset Statistics.}
\label{tab:dataset_stat}
\end{table}

22 Public datasets with varying task/annotator numbers, label sets, label redundancies, and total annotation numbers are collected for the experiments. They can be found from 5 sources~\cite{ipeirotis2010quality,josephy2014workshops,venanzi2015activecrowdtoolkit,zheng2017truth,zhang2014spectral}. We summarize their statistics in Table~\ref{tab:dataset_stat}. Detailed information including average redundancy information and the download links of each dataset can be found in the supplementary material.

In the experiments, we conduct 4-fold splits on every dataset. The size of testing data is 25\% in each fold and the size of validation set is set as 20\% of training data. The results of every experiment are the 4-fold averaged results.

\subsubsection{Baselines}
In our experiments, to provide a comprehensive comparison of previous methods, we use 11 baselines, including majority voting methods MV, WAWA~\cite{crowdkit2023} and ZeroBasedSkill (ZBS)~\cite{crowdkit2023}, probabilistic methods DS~\cite{dawid1979maximum}, GLAD~\cite{whitehill2009whose}, ZC~\cite{demartini2012zencrowd}, MACE~\cite{hovy2013learning}, EBCC~\cite{li2019exploiting} and KOS~\cite{karger2014budget}, neural network based methods LAA~\cite{li2017aggregating} and BiLA~\cite{hong2021online}. Note that, KOS is inherently restricted on binary classification problems and is ill-performing when there is only one annotation for each task. We only report the results of KOS on the datasets having 2 choices.
We also want to clarify that, there are some recent works like SBIC~\cite{manino2019streaming} and LA~\cite{yang2022light} that are not included in the baselines. This is because of that although they made extraordinary theoretical contributions in label aggregation, their performance is not satisfactory enough, where SBIC is nearly the same as or even worse than MV and LA is worse than EBCC according to their original reported results. 
Since the more superior methods like LAA, BiLA and EBCC are included, it will not lead to much more difference in our performance comparisons. 

\subsubsection{Implementation}
LAA, EBCC, and BiLA are implemented based on their original codes. ZC, DS, and GLAD are implemented based on codes from both crowd\_truth\_infer project~\cite{zheng2017truth}. MV, ZBS, WAWA, MACE, and KOS are based on crowdkit~\cite{crowdkit2023}. For our SuperLA, in MLP-1, we use a linear layer with ReLU activation, the hidden size is 16. In MLP-2, we use two linear layers and a dropout layer with a dropout rate of 0.5, the hidden size is set to 8. All embedding sizes for the embedding layer are set to 8. Data replication time in data augmentation is set to 10. L2 regularization is employed with a 0.001 weight decay rate. Early stopping is utilized, monitoring the change of validation loss, with a patience of 5. The overall model is trained by AdamW optimizer, in which the learning rate is set to 0.001, the batch size is 1024. All experiments are conducted on a server with Intel(R) Xeon(R) E5-2680 v4 @ 2.40GHz CPU and one NVIDIA GeForce GTX 1080 GPU.

\subsubsection{Metrics}
Accuracy and F1-score are used to measure the performance. For each dataset, we report the average results among experiments on 4 folds of data, denoted as Acc (F1)\%. Due to the page limitation, we report average performance of each method among all datasets, denoted as Avg. Acc (F1)\%, in the following paragraphs. We also report the count of best Acc (F1)\% of each method, denoted as \#best Acc (F1). The detailed Acc (F1)\% results of each method for each dataset can be referred to the supplementary materials.

\subsection{Performance}

\subsubsection{Main Results}

\begin{table}[htbp]
\centering
\begin{tabular}{lcc}
\hline\hline
Methods &  Avg. Acc (F1)\% & \#best Acc (F1) \\\hline\hline
MV & 79.58$\pm$2.34 (73.28$\pm$2.61) & 2 (1)\\
DS & 81.36$\pm$2.16 (75.84$\pm$2.80) & \underline{3} (\underline{3})\\
GLAD & 81.12$\pm$2.30 (74.30$\pm$2.90) & 3 (2)\\
ZBS & 80.98$\pm$2.21 (74.69$\pm$2.66) & 1 (0)\\
WAWA & 80.71$\pm$2.24 (74.43$\pm$2.65) & 1 (0)\\
ZC & 81.16$\pm$2.41 (74.40$\pm$2.94) & 2 (3)\\
MACE & 80.52$\pm$2.33 (75.40$\pm$2.95) & 2 (1)\\
EBCC & 81.88$\pm$2.15 (75.54$\pm$2.75) & 1 (1)\\
KOS & 80.52$\pm$2.33 (75.40$\pm$2.95) & 0 (0)\\
LAA & 78.45$\pm$2.39 (72.04$\pm$2.95) & 0 (0)\\
BiLA & \underline{82.00$\pm$2.06} (\underline{76.12$\pm$2.62}) & 2 (1)\\\hline
SuperLA & \textbf{85.36$\pm$1.99} (\textbf{79.12$\pm$2.62}) & \textbf{13} (\textbf{11})\\
\hline\hline
\end{tabular}
\caption{Baseline methods are trained on \textit{\textbf{testing}} set. The bold numbers are the best results and the underlined numbers are the second best results.}
\label{tab:performance_test}
\end{table}

Previous baselines can be used in three ways. First, according to their original papers, all baselines are designed to train and test both on testing set. Second, in our 4-fold setting, a more advantageous application strategy of these baselines is to train them on all three datasets, i.e., for each public dataset, we train each baseline method based on the training, validation and testing sets together, and then only reporting the inference results on the testing set.
Third, for methods DS, GLAD and ZC, as introduced in previous literature~\cite{zheng2017truth}, their inference performance can be enhanced by initializing based on historical information. To this end, we compare the performance based on three different utilization strategies of previous baselines: train and test them on testing set, train them on whole dataset and test on testing set, initialize them based on information from training and validation sets and test on testing set.

\textbf{Baselines trained on testing set.} As the most official application strategy of previous methods, we first report the performance comparison under this setting. The averaged accuracy and F1-score across all the datasets are presented in Table~\ref{tab:performance_test}, detailed performance for each dataset can be found in supplementary material. From the results, it is obvious that our SuperLA significantly outperforms all other baselines. On all 22 datasets, SuperLA is ranked 1st on 13 datasets in terms of inference accuracy.

\begin{table}[htbp]
\centering
\begin{tabular}{lcc}
\hline\hline
Methods & Avg. Acc (F1)\% & \#best Acc (F1) \\\hline\hline
MV & 79.58$\pm$2.34 (73.28$\pm$2.61) & 0 (0)\\
DS & 82.88$\pm$2.01 (\underline{77.39$\pm$2.70}) & 3 (3)\\
GLAD & 81.00$\pm$2.45 (73.85$\pm$3.12) & 1 (1)\\
ZBS & 81.42$\pm$2.22 (75.12$\pm$2.72) & 1 (1)\\
WAWA& 81.09$\pm$2.23 (74.79$\pm$2.68) & 0 (0)\\
ZC & 80.26$\pm$2.72 (72.83$\pm$3.34) & 1 (1)\\
MACE & 81.98$\pm$1.98 (76.17$\pm$2.72) & \underline{7} (\underline{5})\\
EBCC & \underline{82.93$\pm$2.14} (76.84$\pm$2.78) & 4 (4)\\
KOS & 77.34$\pm$4.06 (73.42$\pm$4.55) & 0 (0)\\
LAA & 80.53$\pm$2.02 (75.76$\pm$2.66) & 1 (1)\\
BiLA & 78.77$\pm$2.68 (71.28$\pm$3.12) & 0 (0)\\\hline
SuperLA & \textbf{85.36$\pm$1.99} (\textbf{79.11$\pm$2.62}) & \textbf{9} (\textbf{8})\\
\hline\hline
\end{tabular}
\caption{Baseline methods trained on \textit{\textbf{whole}} dataset.}
\label{tab:performance_whole}
\end{table}

\textbf{Baselines trained on whole set.} Under this setting, we allow baselines to use all accessible data to train their model, which is largely beneficial to them. The results are shown in Table~\ref{tab:performance_whole} and detailed results are shown in supplementary. Most of the baselines obtain improvements, while SuperLA still exhibits superior performance and achieves the best method over 9 datasets in terms of accuracy.

\begin{table}[htbp]
\centering
\begin{tabular}{lcc}
\hline\hline
Methods & Avg. Acc (F1) & \#best Acc (F1) \\\hline\hline
DS & 80.53$\pm$2.31 (\underline{75.45$\pm$2.90}) & 3 (3)\\
GLAD & \underline{80.94$\pm$2.31} (74.45$\pm$2.84) & \underline{4} (\underline{5}) \\
ZC & 60.17$\pm$5.75 (54.80$\pm$5.60) & 2 (1)\\\hline
SuperLA & \textbf{85.36$\pm$1.99} (\textbf{79.11$\pm$2.62}) & \textbf{16} (\textbf{13})\\
\hline\hline
\end{tabular}
\caption{Initialize parameters of baseline methods based on training set.}
\label{tab:performance_initial}
\end{table}

\textbf{Initializing baselines based on training and validation set.} We initialize the model parameters of DS, GLAD and ZC methods regarding the calculation of historical information. This is supposed to facilitate the performance of these methods. Results are reported in Table~\ref{tab:performance_initial} and details are referred to the supplementary. Obviously, SuperLA can still obtain the best performance over all other baselines in terms of all three metrics.

The above three results illustrate the inference superiority of SuperLA.

\begin{table}[thbp]
\centering
\begin{tabular}{cccc}
\hline\hline
  A.F & E.F & M.F &  Avg. Acc (F1)\% \\\hline\hline
 \xmark & \cmark & \cmark & 83.34$\pm$2.11 (76.86$\pm$2.81)$\downarrow$ \\
 \cmark & \xmark & \cmark & 84.97$\pm$2.04 (78.31$\pm$2.83)$\downarrow$ \\
 \cmark & \cmark & \xmark & 83.84$\pm$2.23 (76.10$\pm$3.04)$\downarrow$ \\\hline
 \cmark & \cmark & \cmark & \textbf{85.36$\pm$1.99} (\textbf{79.11$\pm$2.62}) \\
\hline\hline
\end{tabular}
\caption{Ablation study. A.F denotes that accuracy feature is removed, E.F denotes that embedded annotator ID feature is removed, and M.F denotes that multi-hot annotator ID feature is removed.}
\label{tab:ablation}
\end{table}

\subsubsection{Ablation}
We ablate each input module of SuperLA to verify the effectiveness of our design. Results are shown in Table~\ref{tab:ablation}. Once there is an absence of any of the designed modules, the performance will be degraded. It indicates that each module can provide useful information that can be automatically learned by SuperLA.

\begin{figure*}
    \centering
    \begin{subfigure}{0.33\textwidth}
        \includegraphics[width=\linewidth]{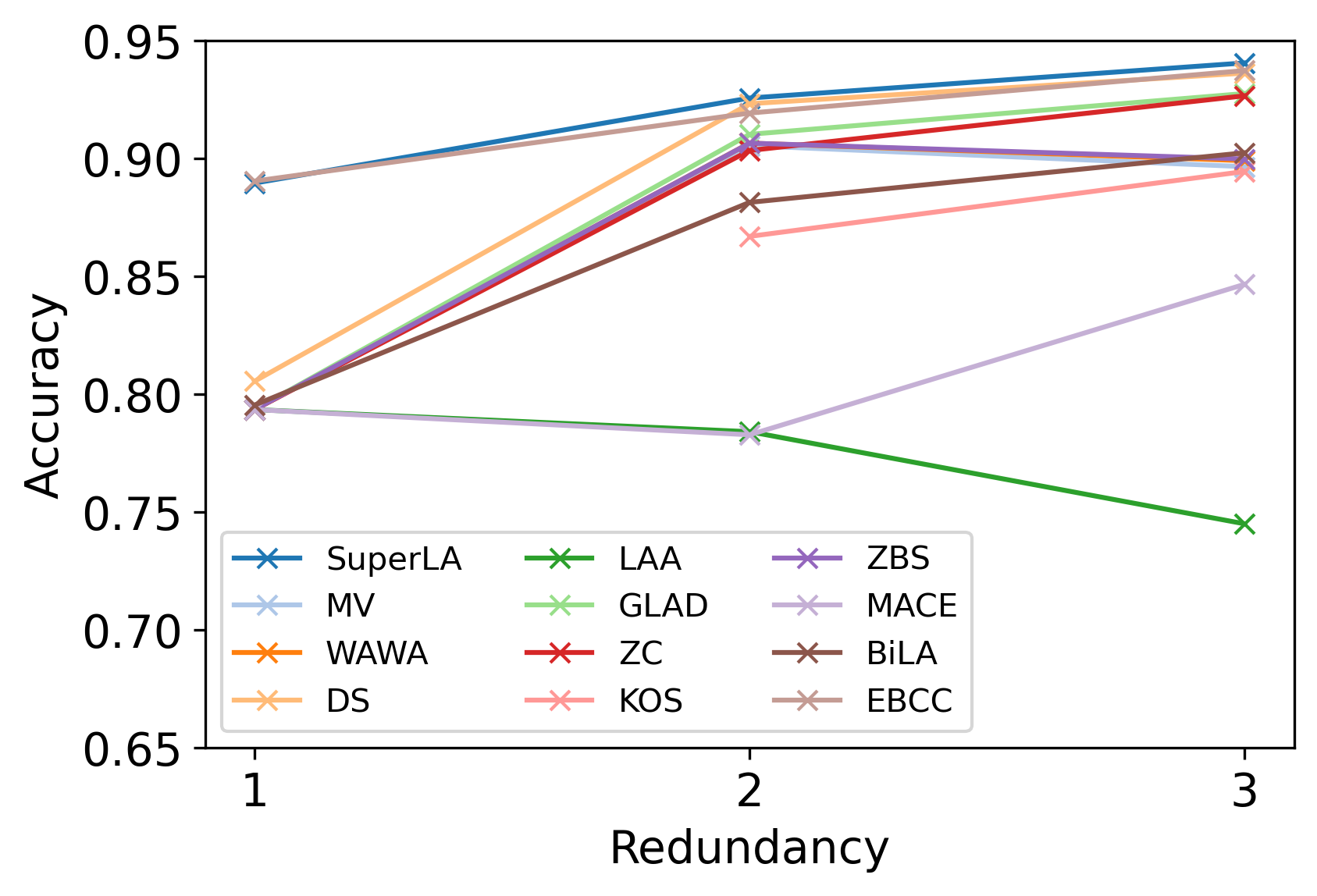}
        \caption{Product, average redundancy 3.}
        \label{fig:redundancy_1}
    \end{subfigure}
    \begin{subfigure}{0.33\textwidth}
        \includegraphics[width=\linewidth]{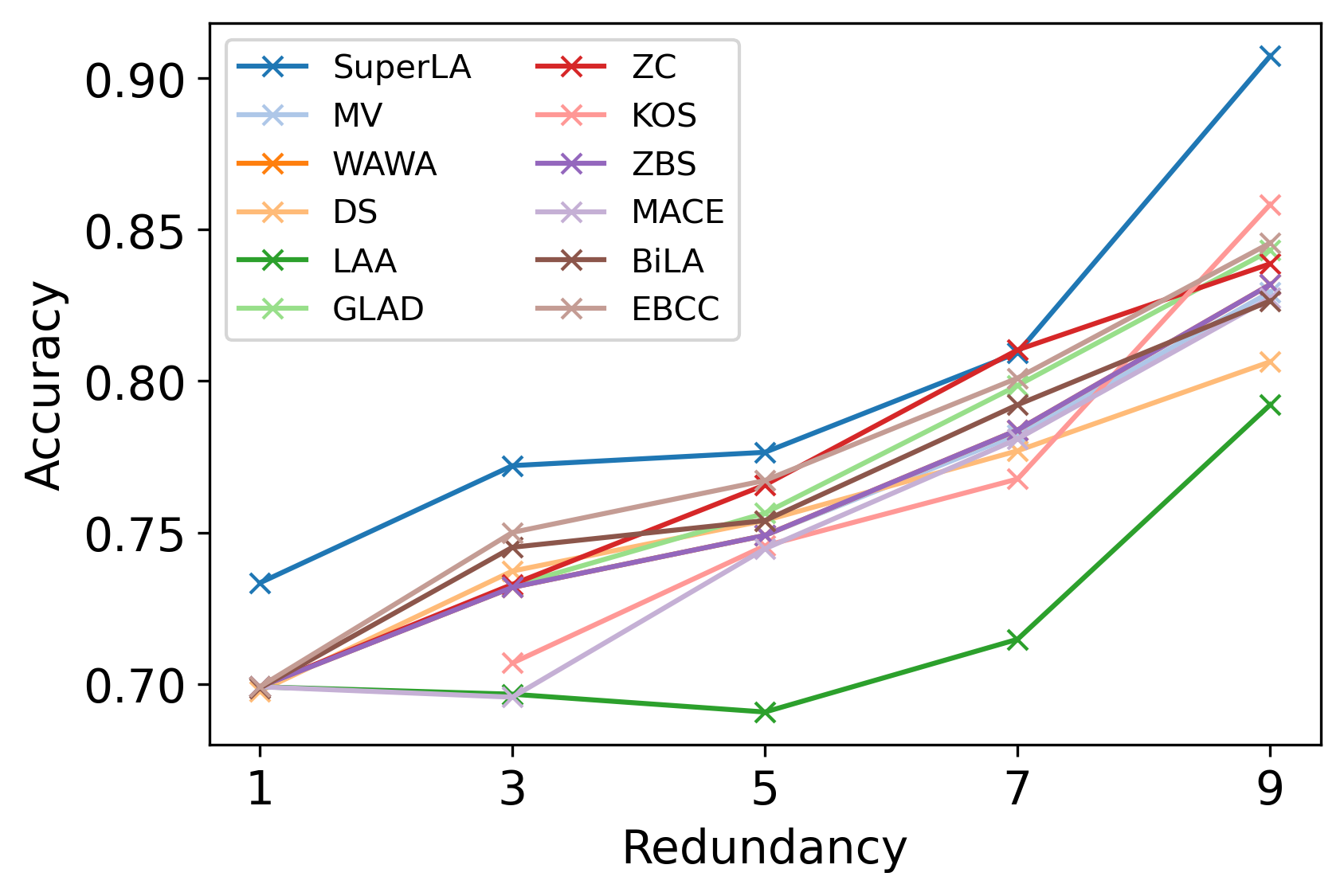}
        \caption{Zencrowd\_all, average redundancy 9.99.}
        \label{fig:redundancy_2}
    \end{subfigure}
    \begin{subfigure}{0.33\textwidth}
        \includegraphics[width=\linewidth]{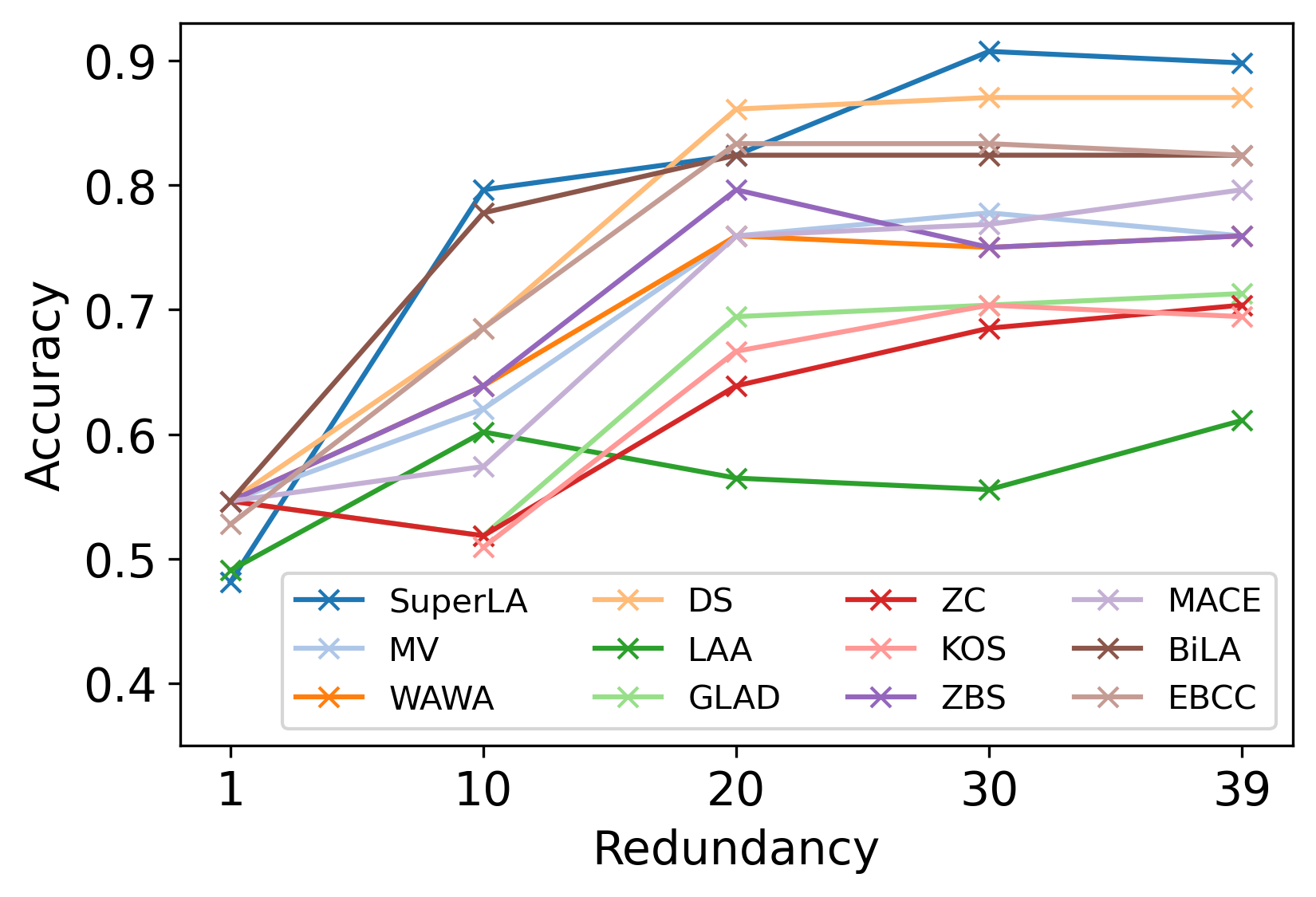}
        \caption{Duck, average redundancy 39.}
        \label{fig:redundancy_2}
    \end{subfigure}
    \caption{Redundancy Analysis.}
    \label{fig:redundancy}
\end{figure*}

\subsubsection{Efficiency}

As it is demonstrated previously, except for trivial methods like MV and WAWA, all other previous works require iterations to solve optimization problems during inference. However, SuperLA only needs optimization on training set. Once trained, no more iteration is needed during inference. We conduct efficient tests over three public datasets that have a relatively large number of tasks to analyze the average inference time on each fold of the testing set. Results are presented in Table~\ref{tab:efficiency}. GLAD has the largest time cost since it has both parameters on annotators and tasks which need substantial time to make the optimization converge. For DS, considering its moderate inference performance and the less time cost compared with other probabilistic methods, it is still an effective and practical method for real-world applications. In any case, SuperLA has very fast inference efficiency. Also, since SuperLA does not need optimization and can be implemented by GPU computation, it can have such a small inference time no matter what size of the target datasets as long as there are enough GPUs. However, other methods will need more inference time when the target size is increasing.

\begin{table}[htbp]
    \centering
    \begin{tabular}{lccc}
    \hline\hline
        \multirow{2}{*}{Methods} & \multicolumn{2}{c}{Datasets} \\\cline{2-4}
         & product & sentiment & relevance \\ \hline\hline
        MV & 0.01956 & 0.016458 & 0.018335 \\ 
        DS & 1.18334 & 0.10995 & 0.86760 \\ 
        GLAD & 36.22101 & 8.95649 & 77.85820 \\
        ZBS & 3.14574 & 2.52294 & 2.92903 \\ 
        WAWA & 0.05206 & 0.03997 & 0.04803 \\ 
        ZC & 0.43109 & 0.15975 & 0.62534 \\
        MACE & 11.32632 & 13.37979 & 14.97524 \\
        EBCC & 4.12855 & 10.30508 & 28.71435 \\ 
        KOS & 1.17791 & - & - \\ 
        LAA & 9.08277 & 2.38086 & 5.57083 \\ 
        BiLA & 3.02842 & 0.53880 & 2.09269 \\\hline
        SuperLA & 0.08272 & 0.06603 & 0.09848 \\ \hline\hline
    \end{tabular}
    \caption{Average inference time (seconds) for testing dataset. SuperLA has a comparable amount of time overhead for inference as trivial methods like MV and WAWA.}
    \label{tab:efficiency}
\end{table}

\begin{figure}[t!]
\centering
\includegraphics[width=0.4\textwidth]{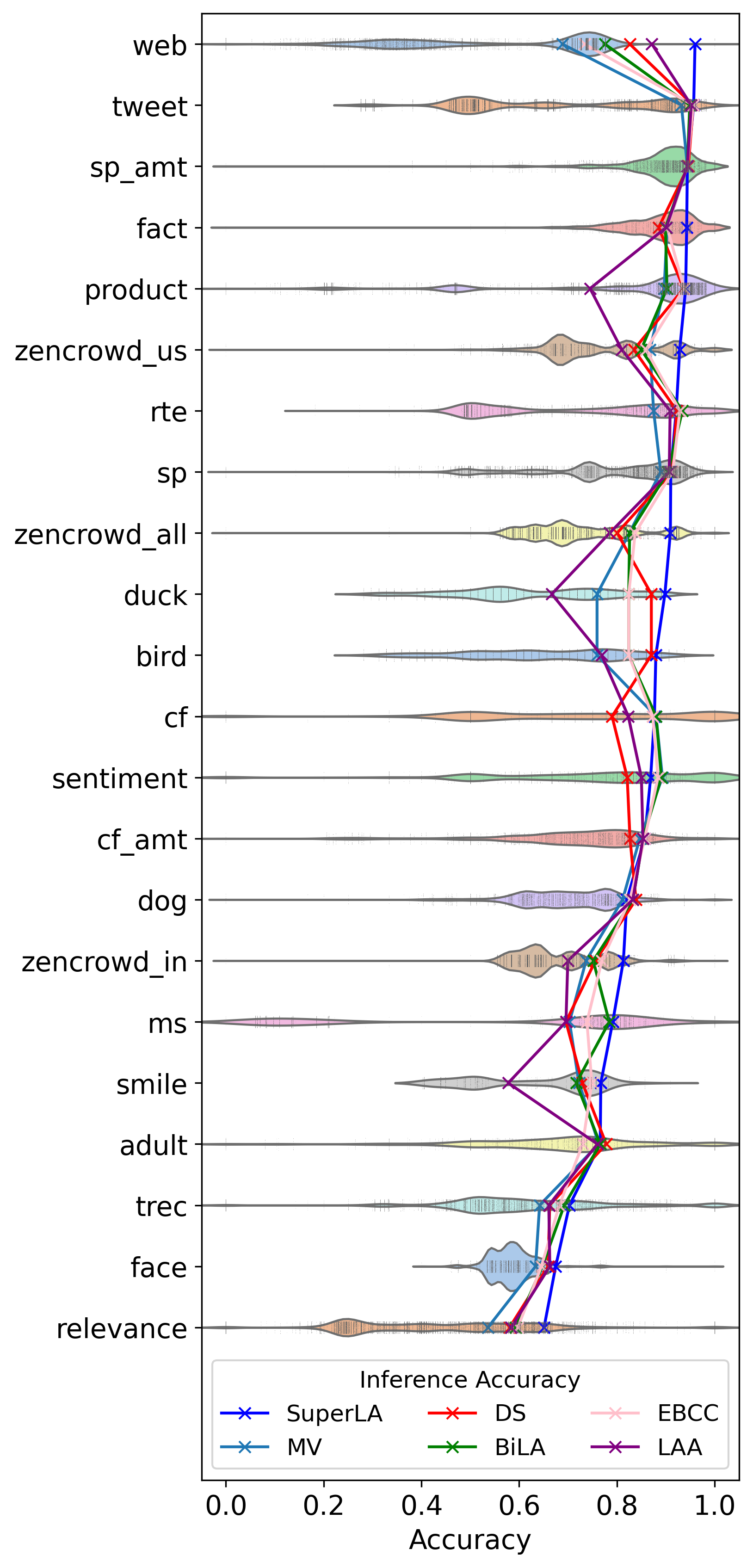} 
\caption{Relations between annotator accuracy distributions and inference accuracies of each dataset.}
\label{fig:dataset_accuracy}
\end{figure}

\subsubsection{Redundancy}
The redundancy is the redundant annotations for each task. With the increasing of redundancy, the cost of the data labeling task will also increase. A higher accuracy in an early stage of labeling that has small redundancy can benefit task assignment or early stopping decisions which finally reduces labeling costs. We analyze the sensitivity to the redundancy of all methods based on case studies of three datasets, product, zencrowd\_all, and duck, which have diverse average redundancy values. For SuperLA, we train it based on full annotations of training set. The performance of all methods are analyzed based on the reduced testing set regarding different redundancies.

As shown in Fig.~\ref{fig:redundancy}, in product and zencrowd\_all, SuperLA obtains the 1st rank accuracy over different redundancies. In the duck dataset, even not the best, SuperLA can obtain comparable inference performance alongside each redundancy step. These results show that, though trained with different redundancy values, superLA can still perform well.


\section{Discussions}

The success of ChatGPT~\cite{ouyang2022training} has ignited the spark for training with large-scale data. The volume of data labeling tasks is increasing faster recently, which meanwhile arises much more demands on label aggregation techniques. However, previous label aggregation techniques are not well prepared for continuous data labeling tasks, especially for the data crowdsourcing platforms that have massive long-term workers and high volume online tasks, where previous techniques will suffer from elusive inference performance and time overhead. Even though some previous works propose to handle this issue to some extent~\cite{feng2014incremental,manino2019streaming,hong2021online,yang2022light}, they make the main efforts to deal with online model updating, which still requires optimizing some object functions and leads to unpredictable converge time. Meanwhile, considering that in real-world applications every update should be carefully investigated and verified before making it functional, so that these incremental model updating techniques still lack practicality.

We point out that to advance label aggregation to a production level, the supervised strategy should be carefully considered. In supervised label aggregation, the training and testing procedures are well separated. Before implementation, we can check the status of the model and monitor the distribution divergence between the pending annotations from the training annotations to have an anticipation of the performance. More importantly, during inferring on current pending tasks, the supervised label aggregation method does not need historical annotations, sequential optimization procedures, or the uncertain inference duration. Contrarily, supervised label aggregation method can infer every task only based on the annotations related to this task, be efficiently paralleled during inference, and has a predictable inference duration. All of these are beneficial to real-world deployment. Based on our experimental results, supervised label aggregation method is able to compete with previous methods in terms of inference performance while having much more advantageous in efficiency.

Furthermore, we provide a relation plot between the inference accuracy of multiple representative methods and the annotator accuracy distributions of each dataset in Fig.~\ref{fig:dataset_accuracy}. Besides that SuperLA achieves a stable good performance on all datasets and the trend of inference accuracy has a generally positive correlation with annotator historical accuracy, there is another interesting finding. When there are fewer annotators having high historical accuracy, other baselines are prone to a decline in inference performance. For example, in datasets web, zencrowd\_us and zencrowd\_all, the annotator accuracy distributions slightly shift to the left and the inference performance of baselines simultaneously drop. However, SuperLA can still obtain relatively good performance in these datasets. This indicates that for SuperLA, it is not always necessary to have annotators with high historical accuracy rates. Using annotators with scattered accuracy can also achieve high label aggregation accuracy. This showcases a possible future direction that to investigate the user accuracy combinations in each task.


\section{Conclusion}
In this paper, we point it out that for the data labeling tasks with long-term annotators, supervised label aggregation is highly desired especially for real-world deployments. We propose SuperLA to form the first method alongside this direction. SuperLA is straightforwardly designed and the structure is not complex, however, it obtains superior inference performance over previous 11 baselines in 22 public datasets and has better inference efficiency and scalability. In the future, we propose to conduct more detailed investigations into possible features and model structures for improving inference performance and will engage in implementing the proposed methods in real-world deployments.

\bibliographystyle{unsrtnat}  
\bibliography{references}

\newpage

\normalsize
\section{Supplementary Materials}

\subsection{Datasets}

\begin{table*}[htb]
\centering
\small
\begin{tabular}{lcccccc}
\hline\hline
Datasets & $|\mathcal{T}|$ & $|\mathcal{A}|$ & $|\mathcal{GT}|$ & $|\mathcal{K}|$ & $|\mathcal{L}|$ & \#Redundancy   \\\hline\hline
adult & 11040 & 825 & 333 & 4 & 89948 & 8.15 \\
bird & 108 & 39 & 108 & 2 & 4212 & 39 \\
CF & 300 & 461 & 300 & 5 & 1720 & 5.73 \\
cf\_amt & 300 & 110 & 300 & 5 & 6025 & 20.08	\\
dog & 807 & 109 & 807 & 4 & 8070 & 10	\\
duck & 108 & 39 & 108 & 2 & 4212 & 39\\
face & 584 & 27 & 584 & 4 & 5242 & 8.98	\\
fact & 42624 & 57 & 576 & 3 & 214960 & 5.04	\\
ms & 700 & 44 & 700 & 10 & 2945 & 4.21	\\
product & 8315 & 176 & 8315 & 2 & 24945 & 3 \\
relevance & 20232 & 766 & 4460 & 4 & 97164 & 4.8\\
rte & 800 & 164 & 800 & 2 & 8000 & 10 \\
sentiment & 98980 & 1960 & 1000 & 5 & 569282 & 5.75 \\
smile & 2134 & 64 & 159 & 2 & 19287 & 9.04 \\
sp & 4999 & 203 & 4999 & 2 & 27746 & 5.55 \\
sp\_amt & 500 & 143 & 500 & 2 & 10000 & 20 \\
trec & 19033 & 762 & 2275 & 2 & 88385 & 4.64 \\
tweet & 1000 & 85 & 1000 & 2 & 20000 & 20.0 \\
web & 2665 & 177 & 2653 & 5 & 15567 & 5.84 \\
zencrowd\_all & 2040 & 78 & 2040 & 2 & 20372 & 9.99 \\
zencrowd\_in & 2040 & 25 & 2040 & 2 & 10626 & 5.21 \\
zencrowd\_us & 2040 & 74 & 2040 & 2 & 11271 & 5.525 \\
\hline\hline
\end{tabular}
\caption{Dataset Statistics.}
\end{table*}

\begin{itemize}
    
    \item Adult dataset is collected by the source~\cite{ipeirotis2010quality} and can be obtained from \url{https://github.com/ipeirotis/Get-Another-Label/tree/master/data}.

    \item Sentiment and fact datasets are collected by CrowdScale2013~\cite{josephy2014workshops} and can be downloaded at \url{https://sites.google. com/site/crowdscale2013/home}.
    
    \item Ms, zencrowd\_all, zencrowd\_us, zencrowd\_in, sp, sp\_amt, cf, and cf\_amt are collected by Active Crowd Toolkit~\cite{venanzi2015activecrowdtoolkit} and can be downloaded at  \url{https://github.com/orchidproject/active-crowd-toolkit}.

    \item Product, tweet, dog, face, duck, relevance and smile are collected from Truth Inference Project~\cite{zheng2017truth} and can be downloaded at \url{https://zhydhkcws.github.io/crowd_truth_inference/datasets.zip}. Note that tweet dataset is called sentiment in this source. It is different from the sentiment dataset in CrowdScale2013.

    \item Bird, rte, web and trec datasets are used in \cite{zhang2014spectral} and can be downloaded at \url{https://github.com/zhangyuc/SpectralMethodsMeetEM}.
    
\end{itemize}

\subsection{Baselines}

\begin{itemize}
    \item MV, ZBS, WAWA, MACE, EBCC, KOS are implemented based on code from Crowdkit~\cite{crowdkit2023}: \url{https://github.com/Toloka/crowd-kit/}.
    \item DS, GLAD, ZC are implemented based on code from previous survey~\cite{zheng2017truth}: \url{https://zhydhkcws.github.io/crowd_truth_inference/index.html}.
    \item LAA~\cite{li2017aggregating} and BiLA~\cite{hong2021online} are implemented based on original code.
\end{itemize}

\subsection{Performance}

Detailed Acc (F1)\% scores that indicate average accuracy and F1-score for 4-fold experiments of each dataset are presented in the following pages.

\begin{table*}[htbp]
\setlength{\abovecaptionskip}{5pt} 
\small
\centering

\begin{tabular}{l|>{\centering\arraybackslash}m{0.9cm}>{\centering\arraybackslash}m{0.9cm}>{\centering\arraybackslash}m{0.9cm}>{\centering\arraybackslash}m{0.9cm}>{\centering\arraybackslash}m{0.9cm}>{\centering\arraybackslash}m{0.9cm}>{\centering\arraybackslash}m{0.9cm}>{\centering\arraybackslash}m{0.9cm}>{\centering\arraybackslash}m{0.9cm}>{\centering\arraybackslash}m{0.9cm}>{\centering\arraybackslash}m{0.9cm}|>{\centering\arraybackslash}m{0.9cm}}
    \hline\hline
        datasets & MV & DS & Glad & ZBS & WAWA & ZC & MACE & EBCC & KOS & LAA & BiLA & SuperLA \\ \hline\hline
        adult & 76.29 (62.25) & \textbf{77.81} (\textbf{67.98}) & 75.69 (61.01) & 76.29 (62.25) & 76.29 (62.25) & 75.99 (61.64) & \underline{76.60} (63.27) & 72.99 (55.81) & - (-) & 76.00 (63.21) & \underline{76.60} (63.04) & 76.58 (\underline{66.04}) \\ \hline
        bird & 75.93 (72.78) & \underline{87.04} (\underline{86.52}) & 78.70 (75.04) & 75.93 (72.78) & 75.93 (72.78) & 78.70 (74.02) & 81.48 (78.92) & 82.41 (81.19) & 68.52 (61.16) & 76.85 (72.71) & 82.41 (81.12) & \textbf{87.96} (\textbf{87.53}) \\ \hline
        cf & \textbf{88.00} (77.17) & 79.00 (70.35) & 87.33 (76.89) & \textbf{88.00} (77.17) & \textbf{88.00} (\underline{77.18}) & 87.33 (\textbf{77.92}) & 86.00 (74.54) & 87.33 (70.80) & - (-) & 82.33 (72.32) & \textbf{88.00} (77.17) & \underline{87.67} (71.12) \\ \hline
        cf\_amt & 84.67 (76.29) & 82.67 (72.65) & \textbf{85.67} (\underline{77.27}) & 85.00 (76.59) & 85.00 (76.59) & \textbf{85.67} (74.01) & \textbf{85.67} (74.01) & \textbf{85.67} (\underline{77.09}) & - (-) & \underline{85.33} (74.05) & \underline{85.33} (76.83) & \textbf{85.67} (73.57) \\ \hline
        dog & 81.16 (80.67) & \textbf{83.89} (\textbf{83.85}) & 82.89 (82.59) & 82.27 (81.90) & 82.27 (81.90) & 82.77 (82.54) & 82.89 (82.65) & 82.65 (82.52) & - (-) & 83.26 (\underline{83.11}) & \underline{83.27} (83.08) & 82.03 (82.00) \\ \hline
        duck & 75.93 (73.97) & \underline{87.04} (\underline{86.80}) & 71.30 (67.54) & 75.93 (73.97) & 75.93 (73.97) & 70.37 (66.27) & 79.63 (77.90) & 82.41 (81.47) & 69.44 (63.81) & 66.67 (57.43) & 82.41 (81.55) & \textbf{89.81} (\textbf{89.68}) \\ \hline
        face & 63.36 (61.34) & 65.92 (64.24) & 63.01 (60.87) & 63.18 (61.18) & 63.01 (60.96) & 63.01 (60.95) & 63.36 (61.06) & 64.55 (62.48) & - (-) & \underline{66.27} (\underline{64.68}) & 64.73 (62.75) & \textbf{67.47} (\textbf{66.71}) \\ \hline
        fact & \underline{90.28} (\underline{45.76}) & 88.54 (44.12) & 90.10 (45.56) & \underline{90.28} (\underline{45.76}) & \underline{90.28} (\underline{45.76}) & 90.10 (45.56) & \underline{90.28} (45.71) & \underline{90.28} (45.71) & - (-) & 90.10 (45.65) & 89.93 (45.38) & \textbf{94.27} (\textbf{52.84}) \\ \hline
        ms & 70.29 (69.63) & 69.57 (68.49) & 79.29 (79.15) & 79.00 (79.02) & 79.00 (79.02) & \underline{79.43} (\underline{79.58}) & \textbf{80.00} (\textbf{80.08}) & 73.86 (72.85) & - (-) & 69.57 (68.11) & 78.43 (78.36) & 79.14 (78.77) \\ \hline
        product & 89.66 (76.56) & 93.64 (83.65) & 92.72 (79.58) & 89.98 (77.04) & 89.91 (76.94) & 92.66 (79.48) & 84.67 (72.93) & \underline{93.73} (\underline{83.78}) & 89.45 (71.84) & 74.50 (62.96) & 90.23 (78.52) & \textbf{94.06} (\textbf{84.52}) \\ \hline
        relevance & 53.61 (53.52) & 58.04 (55.06) & 58.27 (56.64) & 56.39 (55.82) & 55.87 (55.43) & 53.43 (52.88) & 51.26 (50.72) & \underline{59.73} (\underline{57.71}) & - (-) & 58.36 (53.89) & 59.26 (57.05) & \textbf{65.09} (\textbf{62.20}) \\ \hline
        rte & 87.50 (87.22) & 92.25 (92.17) & 92.50 (92.41) & 90.63 (90.49) & 89.88 (89.69) & 91.75 (91.67) & 92.75 (92.67) & \underline{92.88} (\underline{92.79}) & 54.62 (47.6) & 90.88 (90.81) & \textbf{93.38} (\textbf{93.29}) & 92.13 (92.04) \\ \hline
        sentiment & \textbf{89.30} (\textbf{77.27}) & 82.10 (69.63) & 87.90 (75.12) & \underline{89.10} (\underline{77.07}) & \underline{89.10} (77.06) & 88.10 (75.51) & 87.30 (74.24) & 88.40 (75.22) & - (-) & 85.00 (74.08) & \underline{89.10} (76.90) & 86.90 (70.38) \\ \hline
        smile & 72.23 (70.56) & 72.95 (70.48) & 75.40 (69.39) & 72.93 (70.78) & 72.29 (70.2) & 75.40 (69.37) & \underline{75.43} (\underline{72.63}) & 74.86 (\textbf{73.04}) & 74.17 (69.91) & 57.76 (53.31) & 71.67 (69.23) & \textbf{76.70} (72.23) \\ \hline
        sp & 88.96 (88.96) & 91.30 (91.30) & \textbf{91.60} (\textbf{91.59}) & 89.48 (89.47) & 89.50 (89.49) & 91.42 (91.41) & \underline{91.46} (\underline{91.45}) & 91.40 (91.39) & 81.92 (81.86) & 90.70 (90.69) & 90.46 (90.45) & 91.02 (91.01) \\ \hline
        sp\_amt & \underline{94.40} (\underline{94.39}) & \textbf{94.60} (\textbf{94.59}) & \textbf{94.60} (\textbf{94.59}) & \underline{94.40} (\underline{94.39}) & \underline{94.40} (\underline{94.39}) & \textbf{94.60} (\textbf{94.59}) & 94.20 (94.19) & \underline{94.40} (\underline{94.39}) & 94.20 (94.19) & \underline{94.40} (\underline{94.39}) & \underline{94.40} (\underline{94.39}) & \underline{94.40} (94.38) \\ \hline
        trec & 64.18 (59.16) & 66.24 (64.55) & 60.26 (49.62) & 64.26 (59.24) & 64.22 (59.23) & 60.31 (50.86) & 61.93 (54.23) & 68.22 (65.06) & 56.17 (43.74) & 66.06 (65.61) & \underline{69.23} (\underline{67.60}) & \textbf{70.29} (\textbf{69.67}) \\ \hline
        tweet & 93.20 (93.14) & \underline{95.60} (\underline{95.56}) & 94.50 (94.43) & 94.80 (94.75) & 94.60 (94.54) & 94.90 (94.84) & 95.50 (95.45) & \underline{95.60} (\underline{95.56}) & 93.80 (93.70) & 95.20 (95.15) & 94.90 (94.84) & \textbf{95.70} (\textbf{95.67}) \\ \hline
        web & 68.83 (68.96) & 82.62 (82.30) & 78.55 (78.74) & 79.23 (79.39) & 75.65 (75.87) & 83.90 (84.11) & 83.07 (83.15) & 73.39 (73.09) & - (-) & \underline{87.15} (\underline{86.82}) & 77.53 (77.45) & \textbf{95.97} (\textbf{96.16}) \\ \hline
        zencrowd\_all & 82.55 (76.62) & 80.05 (75.55) & 82.65 (77.71) & 83.48 (77.36) & 83.28 (77.16) & 82.79 (77.85) & 81.08 (76.25) & 83.73 (78.65) & \underline{85.34} (\underline{80.40}) & 78.53 (74.22) & 82.60 (77.61) & \textbf{90.93} (\textbf{86.02}) \\ \hline
        zencrowd\_in & 73.68 (65.76) & 75.59 (69.65) & 75.05 (67.23) & 74.46 (66.32) & 74.46 (66.32) & \underline{76.96} (\textbf{70.83}) & 74.26 (\underline{69.0}) & 76.91 (69.89) & 75.15 (68.84) & 69.95 (65.11) & 75.15 (68.02) & \textbf{81.32} (68.94) \\ \hline
        zencrowd\_us & 86.76 (80.08) & 83.48 (78.89) & 86.76 (81.61) & 86.52 (80.54) & 86.76 (80.72) & 85.83 (80.88) & 84.07 (79.64) & 85.93 (81.39) & \underline{87.35} (\underline{81.91}) & 80.98 (76.56) & 84.95 (80.12) & \textbf{92.84} (\textbf{88.86}) \\ \hline\hline
        average & 79.58 (73.28) & 81.36 (75.84) & 81.12 (74.30) & 80.98 (74.69) & 80.71 (74.43) & 81.16 (74.40) & 80.52 (75.40) & 81.88 (75.54) & 77.51 (71.58) & 78.45 (72.04) & \underline{82.00} (\underline{76.12}) & \textbf{85.36} (\textbf{79.11})\\\hline
    \hline
\end{tabular}
\caption{Baseline methods train on \textit{\textbf{testing}} dataset. Acc (F1)\% results.}
\end{table*}

\begin{table*}[htbp]
\setlength{\abovecaptionskip}{5pt} 
\small
\centering

\begin{tabular}{l|>{\centering\arraybackslash}m{0.9cm}>{\centering\arraybackslash}m{0.9cm}>{\centering\arraybackslash}m{0.9cm}>{\centering\arraybackslash}m{0.9cm}>{\centering\arraybackslash}m{0.9cm}>{\centering\arraybackslash}m{0.9cm}>{\centering\arraybackslash}m{0.9cm}>{\centering\arraybackslash}m{0.9cm}>{\centering\arraybackslash}m{0.9cm}>{\centering\arraybackslash}m{0.9cm}>{\centering\arraybackslash}m{0.9cm}|>{\centering\arraybackslash}m{0.9cm}}
    \hline\hline
        datasets & MV & DS & Glad & ZBS & WAWA & ZC & MACE & EBCC & KOS & LAA & BiLA & SuperLA \\ \hline\hline
        adult* & 76.29 (62.25) & 76.58 (\textbf{66.22}) & 75.99 (61.21) & \underline{76.60} (62.47) & \underline{76.60} (62.47) & 72.09 (56.77) & \textbf{76.90} (63.17) & 74.79 (61.86) & - (-) & 76.59 (65.69) & 73.88 (\underline{66.14}) & 76.58 (66.04) \\ \hline
        bird & 75.93 (72.78) & \textbf{88.89} (\textbf{88.40}) & 72.22 (66.35) & 75.93 (72.78) & 75.93 (72.78) & 72.22 (66.35) & 86.11 (85.26) & 86.11 (85.26) & 72.22 (66.35) & 76.85 (72.6) & 57.41 (42.83) & \underline{87.96} (\underline{87.53}) \\ \hline
        cf & 88.00 (77.17) & 83.00 (73.34) & 88.00 (77.17) & 88.00 (\underline{77.24}) & 88.00 (\underline{77.24}) & 88.00 (75.1) & \textbf{88.67} (75.66) & \underline{88.33} (71.59) & - (-) & 86.00 (\textbf{78.56}) & 86.67 (70.58) & 87.67 (71.12) \\ \hline
        cf\_amt & 84.67 (76.29) & 84.00 (73.05) & 85.33 (76.89) & 85.67 (\underline{77.13}) & 85.67 (\underline{77.13}) & 85.33 (73.63) & 85.67 (74.00) & \textbf{87.00} (74.89) & - (-) & \underline{86.00} (\textbf{77.31}) & 82.67 (66.48) & 85.67 (73.57) \\ \hline
        dog & 81.16 (80.67) & \textbf{84.26} (\textbf{84.31}) & 83.39 (83.14) & 82.89 (82.60) & 82.89 (82.60) & 83.02 (82.83) & 83.02 (82.87) & 84.01 (83.97) & - (-) & \underline{84.13} (\underline{84.12}) & 82.03 (81.73) & 82.03 (82.00) \\ \hline
        duck & 75.93 (73.97) & \underline{88.89} (\underline{88.68}) & 72.22 (67.73) & 75.93 (73.97) & 75.93 (73.97) & 72.22 (67.73) & 86.11 (85.65) & 86.11 (85.55) & 72.22 (67.73) & 86.11 (85.70) & 87.96 (84.32) & \textbf{89.81} (\textbf{89.68}) \\ \hline
        face & 63.36 (61.34) & 64.00 (62.31) & 63.01 (60.88) & 63.01 (60.91) & 63.01 (60.91) & 62.84 (60.69) & 64.38 (62.21) & 64.38 (62.11) & - (-) & \underline{66.27} (\underline{64.75}) & 64.04 (61.78) & \textbf{67.47} (\textbf{66.71}) \\ \hline
        fact & 90.28 (45.76) & 89.24 (44.71) & 90.28 (45.71) & 90.10 (45.56) & 90.10 (45.56) & 90.10 (45.56) & 87.50 (43.28) & 88.37 (44.11) & - (-) & 83.16 (\underline{48.63}) & \underline{90.80} (42.71) & \textbf{94.27} (\textbf{52.84}) \\ \hline
        ms & 70.29 (69.63) & 76.86 (76.42) & 78.86 (78.72) & \underline{79.86} (\underline{79.87}) & 79.71 (79.77) & 79.57 (79.72) & \textbf{80.00} (\textbf{80.09}) & 78.71 (78.48) & - (-) & 78.29 (78.27) & 69.71 (69.86) & 79.14 (78.77) \\ \hline
        product & 89.66 (76.56) & \underline{93.96} (\underline{84.33}) & 92.82 (79.81) & 89.66 (76.56) & 89.66 (76.56) & 92.80 (79.81) & 81.62 (70.16) & 93.49 (83.87) & 89.55 (72.26) & 73.83 (63.01) & 89.55 (77.7) & \textbf{94.05} (\textbf{84.52}) \\ \hline
        relevance* & 53.61 (53.52) & \underline{61.73} (\underline{59.57}) & 56.46 (55.77) & 56.01 (55.59) & 56.59 (55.98) & 46.57 (46.33) & 60.78 (57.92) & 59.73 (58.49) & - (-) & 58.27 (55.24) & 45.61 (48.65) & \textbf{65.09} (\textbf{62.20}) \\ \hline
        rte & 87.50 (87.22) & \underline{92.75} (\underline{92.67}) & 92.50 (92.42) & 92.13 (92.03) & 91.88 (91.77) & 92.50 (92.42) & 92.63 (92.54) & \textbf{93.12} (\textbf{93.05}) & 49.75 (42.55) & 92.00 (91.92) & 81.13 (80.64) & 92.13 (92.04) \\ \hline
        sentiment & \underline{89.30} (\underline{77.27}) & 87.40 (76.71) & \underline{89.30} (77.23) & 89.10 (77.05) & \underline{89.30} (77.21) & 88.90 (76.94) & \textbf{89.50} (\textbf{78.32}) & 86.00 (73.19) & - (-) & 80.60 (73.13) & 87.10 (74.18) & 86.90 (70.38) \\ \hline
        smile & 72.23 (70.56) & \textbf{77.95} (\textbf{74.82}) & 72.29 (62.96) & 76.07 (\underline{73.76}) & 74.18 (71.98) & 72.29 (62.96) & 74.20 (71.64) & 72.32 (70.56) & 70.42 (59.15) & 73.56 (71.56) & 74.21 (72.18) & \underline{76.70} (72.23) \\ \hline
        sp & 88.96 (88.96) & 91.52 (91.51) & \underline{91.68} (\underline{91.67}) & 89.46 (89.45) & 89.50 (89.49) & 91.66 (91.65) & \textbf{91.78} (\textbf{91.77}) & 91.26 (91.25) & 81.92 (81.86) & 91.54 (91.53) & 89.78 (89.77) & 91.02 (91.01) \\ \hline
        sp\_amt & \underline{94.40} (\underline{94.39}) & \underline{94.40} (\underline{94.39}) & \textbf{94.60} (\textbf{94.59}) & \textbf{94.60} (\textbf{94.59}) & \underline{94.40} (\underline{94.39}) & \textbf{94.60} (\textbf{94.59}) & \textbf{94.60} (\textbf{94.59}) & \textbf{94.60} (\textbf{94.59}) & 94.20 (94.19) & \underline{94.40} (\underline{94.39}) & 91.80 (91.14) & \underline{94.40} (94.38) \\ \hline
        trec & 64.18 (59.16) & 70.15 (68.43) & 57.93 (43.17) & 64.44 (59.2) & 64.13 (59.12) & 57.01 (41.06) & \textbf{70.99} (\textbf{70.35}) & \underline{70.37} (68.44) & 56.64 (43.00) & 68.75 (68.27) & 66.99 (54.18) & 70.29 (\underline{69.67}) \\ \hline
        tweet & 93.20 (93.14) & \underline{96.00} (\underline{95.96}) & 94.80 (94.73) & 95.30 (95.25) & 95.10 (95.05) & 95.10 (95.04) & 95.70 (95.65) & \textbf{96.10} (\textbf{96.07}) & 93.30 (93.20) & 95.60 (95.56) & 94.30 (94.28) & 95.70 (95.67) \\ \hline
        web & 68.83 (68.96) & 82.92 (82.54) & 82.13 (82.48) & 81.87 (82.22) & 76.93 (77.17) & 83.98 (84.26) & 80.70 (80.85) & 74.37 (74.28) & - (-) & \underline{87.94} (\underline{87.42}) & 69.51 (69.43) & \textbf{95.97} (\textbf{96.16}) \\\hline
        zencrowd\_all & 82.55 (76.62) & 80.25 (75.82) & 83.04 (78.16) & 83.73 (77.65) & 83.28 (77.13) & 82.79 (77.77) & 79.12 (74.88) & \underline{86.23} (\underline{81.08}) & 85.44 (80.52) & 78.53 (74.37) & 84.85 (78.63) & \textbf{90.93} (\textbf{86.02}) \\ \hline
        zencrowd\_in & 73.68 (65.76) & 75.64 (70.05) & 77.16 (\underline{70.87}) & 74.51 (66.36) & 74.56 (66.45) & 76.81 (70.62) & 72.75 (68.31) & \underline{77.79} (\textbf{71.06}) & 74.85 (68.28) & 72.16 (67.75) & 74.71 (67.24) & \textbf{81.32} (68.94) \\ \hline
        zencrowd\_us* & 86.76 (80.08) & 82.94 (78.39) & 87.89 (83.09) & 86.32 (80.36) & 86.62 (80.56) & 85.39 (80.48) & 80.83 (76.60) & \underline{91.23} (\underline{86.82}) & 87.59 (81.58) & 81.18 (76.94) & 88.28 (83.74) & \textbf{92.84} (\textbf{88.86}) \\ \hline
        average & 79.58 (73.28) & 82.89 (\underline{77.39}) & 81.00 (73.85) & 81.42 (75.12) & 81.09 (74.79) & 80.26 (72.83) & 81.98 (76.17) & \underline{82.93} (76.84) & 77.34 (73.42) & 80.53 (75.76) & 78.77 (71.28) & \textbf{85.36} (\textbf{79.11}) \\\hline
    \hline
\end{tabular}
\caption{Baseline methods train on \textbf{\textit{whole}}. Acc (F1)\% results. } 
\end{table*}

\begin{table*}[htbp]
\setlength{\abovecaptionskip}{5pt} 
\small
\centering

\begin{tabular}{l|>{\centering\arraybackslash}m{0.9cm}>{\centering\arraybackslash}m{0.9cm}>{\centering\arraybackslash}m{0.9cm}|>{\centering\arraybackslash}m{0.9cm}}
    \hline\hline
    datasets & DS & GLAD & ZC & SuperLA \\ \hline
        adult & 73.88 (63.63) & 75.69 (60.56) & \underline{76.29} (\underline{64.11}) & \textbf{76.58} (\textbf{\textbf{66.04}}) \\ \hline
        duck & \underline{88.89} (\underline{88.68}) & 74.07 (70.66) & 29.63 (24.26) & \textbf{89.81} (\textbf{89.68}) \\ \hline
        face & \underline{66.78} (\underline{65.55}) & 62.84 (60.66) & 63.01 (60.95) & \textbf{67.47} (\textbf{66.71}) \\ \hline
        relevance & \underline{58.65} (\underline{55.53}) & 55.56 (54.81) & 53.70 (53.10) & \textbf{65.09} (\textbf{62.20}) \\ \hline
        rte & \underline{91.88} (\underline{91.8}) & 91.25 (91.16) & 30.25 (30.18) & \textbf{92.13} (\textbf{92.04}) \\ \hline
        sentiment & 74.20 (64.12) & \textbf{88.90} (\textbf{76.61}) & \underline{88.40} (\underline{75.57}) & 86.90 (70.38) \\ \hline
        product & \underline{93.64} (\underline{83.74}) & 92.06 (77.45) & 7.44 (7.26) & \textbf{94.05} (\textbf{84.52}) \\ \hline
        bird & \textbf{88.89} (\textbf{88.40}) & 78.7 (75.04) & 26.85 (20.82) & \underline{87.96} (\underline{87.53}) \\ \hline
        cf & 66.33 (56.04) & \textbf{88.00} (\textbf{78.90}) & 86.67 (\underline{77.32}) & \underline{87.67} (71.12) \\ \hline
        cf\_amt & \underline{84.67} (\underline{75.49}) & \textbf{85.67} (\textbf{77.27}) & \textbf{85.67} (74.01) & \textbf{85.67} (73.57) \\ \hline
        dog & \textbf{84.50} (\textbf{84.51}) & \underline{83.14} (\underline{82.84}) & 82.77 (82.54) & 82.03 (82.00) \\ \hline
        fact & 87.85 (\underline{49.61}) & \underline{89.76} (45.22) & 68.06 (35.18) & \textbf{94.27} (\textbf{52.84}) \\ \hline
        MS & 68.14 (66.94) & 78.71 (78.46) & \textbf{79.43} (\textbf{79.58}) & \underline{79.14} (\underline{78.77}) \\ \hline
        smile & \underline{74.20} (\underline{71.64}) & \underline{74.20} (69.83) & 45.42 (42.26) & \textbf{76.70} (\textbf{72.23}) \\ \hline
        sp & 91.30 (91.30) & \textbf{91.68} (\textbf{91.67}) & \underline{91.38} (\underline{91.37}) & 91.02 (91.01) \\ \hline
        sp\_amt & \textbf{94.40} (\textbf{94.39}) & \underline{94.20} (94.19) & 74.4 (74.39) & \textbf{94.40} (\underline{94.38}) \\ \hline
        trec & \underline{66.11} (\underline{65.66}) & 61.32 (52.44) & 57.63 (47.37) & \textbf{70.29} (\textbf{69.67}) \\ \hline
        tweet & \underline{95.60} (\underline{95.56}) & 95.20 (95.15) & 5.20 (5.05) & \textbf{95.70} (\textbf{95.67}) \\ \hline
        zencrowd\_all & 79.26 (74.87) & 81.37 (76.65) & \underline{82.75} (\underline{77.80}) & \textbf{90.93} (\textbf{86.02}) \\ \hline
        zencrowd\_in & 75.34 (\underline{69.83}) & \underline{76.47} (\textbf{70.88}) & 36.37 (33.74) & \textbf{81.32} (68.94) \\ \hline
        zencrowd\_us & 82.25 (77.89) & \underline{83.92} (\underline{79.28}) & 68.38 (64.49) & \textbf{92.84} (\textbf{88.86}) \\ \hline
        web & \underline{84.96} (\underline{84.73}) & 77.95 (78.10) & 83.94 (84.17) & \textbf{95.97} (\textbf{96.16}) \\\hline\hline
        average & 80.53 (75.45) & 80.94 (74.45) & 60.17 (54.80) & 85.36 (79.11) \\\hline
    \hline
\end{tabular}
\caption{Initialize parameters of baseline methods based on training dataset. Acc (F1)\% results.}
\end{table*}

\end{document}